# Optimization of the image contrast in SPECT-CT bremsstrahlung imaging for Selective Internal Radiation Therapy of liver malignancies with Y-90 microspheres

F. Bonutti[1], M. Avolio[2], G. Magro[3,4], A. Cecotti[5], E. Della Schiava[1], E. Del Dò[1], F. Longo[6], M.Y. Herassi[7], F. Bentayeb[7], M. Rossi[5], G. Ferretti[5], O. Geatti[5] and R. Padovani[8]

[1] Academic Hospital of Udine, Medical Physics Department, Italy

[2] University of Trieste, Department of Physics, Italy

[3] University of Pavia, Department of Physics, Italy

[4] CNAO, National Center for Oncological Hadron Therapy, Pavia, Italy

[5] Academic Hospital of Udine, Nuclear Medicine Department, Italy

[6] INFN, Istituto Nazionale di Fisica Nucleare, sezione di Trieste, Italy

[7] Faculty of Sciences of Rabat, (LPHE, MS), Department of Physics, Morocco

[8] International Master of Medical Physics, ICTP, Trieste, Italy

Corresponding author:
Faustino Bonutti
Medical Physics Department, Academic Hospital of Udine, Italy
Piazzale Santa Maria della Misericordia, 15
33100 Udine - ITALY
tel. ++39 0432 554574
fax. ++39 0432 552548
email: bonutti.faustino@aoud.sanita.fvg.it

**Abstract**

The quality of SPECT Bremsstrahlung images of patients treated with Y-90 is poor, mainly because of scattered radiation and collimator septa penetration. To minimize the latter effect, High Energy (HE) or Medium Energy (ME) collimators can be used. Scatter correction is not possible through the methods commonly used for the diagnostic radionuclides (Tc-99m, etc.) because the Bremsstrahlung radiation does not have distinct photopeaks, but a broad spectrum of energies ranging from zero to the maximum one detectable by the gamma-camera crystal is registered. Scatter radiation and collimator septa penetration affect the Contrast and the Contrast Recovery Coefficient (CRC) : our research focused on finding the best energy position for the acquisition window in order to maximize these parameters. To be guided in this finding, we first made a Monte Carlo (MC) simulation of a SPECT acquisition of a Y-90 cylindrical phantom and then we measured at different energies the Line Spread Function (LSF) of a linear Y-90 source inserted in a scatter medium. Finally we acquired the NEMA IEC Body phantom filled with Y-90 chloride at different energies and using different setups in order to measure the Contrast and the CRC in different conditions with the purpose of optimizing the clinical SPECT acquisition procedures. The final results showed that Contrast and CRC are higher for HE collimators compared to ME collimators and they have a maximum for the larger spheres at ~ 110 - 135 keV.

**Introduction**

In a recent work **[1],** *Elschot, M. et al.* demonstrated the higher image quality of the state-of-the-art PET compared to Bremsstrahlung SPECT in imaging of in-vivo Y-90 microsphere distribution after liver radio-embolization. If a PET-CT scanner is not available in a Nuclear Medicine Department and the SPECT-CT is used to verify the microspheres distributions, then the acquisitions need to be optimized in order to improve the image quality as much as possible. The main difficulty of SPECT imaging of patients treated with Y-90 microspheres is due to the fact that most of the Bremsstrahlung radiation (primary photons), produced via the interaction of the $\beta^-$s with the patient's tissues, undergoes Compton scattering and therefore loses its original direction. The energy spectrum of photons emerging from the patient and measured by a gamma-camera detector has no photopeaks and the traditional methods for scatter radiation correction are not applicable. Furthermore, the SPECT detectors are designed to acquire photons emitted by the conventional diagnostic radionuclides which have energies up to 364 keV (I-131) or at most up to 511 keV (F-18), whilst the maximum energy of the Bremsstrahlung radiation is equal to the Y-90 $\beta^-$ maximum energy $E_{max} = 2.28$ MeV. Finally, photons belonging to the highest energy region of the Bremsstrahlung spectrum pass through the collimator septa, lowering the image quality. Several studies have been undertaken to bring improvement to the situation. In a previous study [2], a reconstruction method of SPECT Y-90, based on a Monte Carlo (MC) simulator adopted for Y-90 and incorporated into a statistical reconstruction algorithm (SPECTMC), was developed. The attenuation and scattered effects were modeled during the reconstruction of Monte Carlo. A phantom experiment was also performed. The authors showed that the method proposed has improved quantitatively the accuracy of the images of Y-90 Bremsstrahlung spectrum with respect to the clinical reconstruction. Based on a dosimetric evaluation of a number of patients, the authors concluded that the proposed method can be used as an alternative to Y-90 PET.

Besides choosing the most appropriate collimators, a way to minimize as much as possible the impact of these effects on the image quality without operating any corrections on the acquired raw data, is to optimize the energy position of the acquisition window.

To carry out the present work, two aspects were considered:

**1.** MC simulation of a SPECT acquisition of a Y-90 cylindrical phantom, using the Geant4/Gate toolkit [3] to model in the interaction of particles with matter.

**2.** Experimental parts: a Body phantom (mod. PET/IEC-Body/P) filled with Y-90 was acquired to measure the image quality parameters. In addition, a linear source of Y-90 was used to measure the Line Spread Function (LSF).

A substantial part of the simulation work consisted of testing and analyzing the capabilities of the MC code, by using two different models for the Electromagnetic (EM) Physics, in order to find the best setup to reproduce the real experiment.
As the higher energy portion of the Bremsstrahlung spectrum is generally characterized by a higher fraction of primary photons, the effect of septa penetration increases with the energy. A measure of the LSF of a linear Y-90 source inserted in a scattering medium was needed to have information about the influence of these two opposed effects on the LSF shape.
Moreover, to measure the Contrast and the CRC at different conditions with the objective of optimizing the clinical SPECT acquisition procedures, the phantom mod. PET/IEC-Body/P filled with Y-90 chloride at different energies and using different setups was acquired.

**Materials and methods**

When a gamma-camera is used for imaging with a single photon emitting radionuclide, the energy window is centered on the photopeak determining the choice of the collimator. In the case of Bremsstrahlung imaging, a distinct photopeak is not present. Instead, a broad spectrum of energies ranging from zero to the maximum that can be detected by the crystal is registered. The energy window settings represent a problem which must be addressed if SPECT-CT is used to assess the Y-90 microspheres distribution inside the liver. When photon energies are enough to penetrate the collimator septa, spatial resolution becomes poorer. It is well established that HE and ME collimators gives the best Contrast and spatial resolution in Y-90 imaging [4].

Bremsstrahlung production by $\beta^-$ particles interaction in low-Z materials is a low-probability event. Sensitivity of Bremsstrahlung imaging will therefore be very small, since less than 1% of the kinetic energy of the $\beta^-$ particle is emitted as Bremsstrahlung radiation for Y-90, and this is a limitation on the expected image quality. To overcome this problem, i.e. to increase the counting statistics, a large window could be used, but it includes a large contribution of scattered radiation. Due to the complexity of the continuous spectrum of the Bremsstrahlung radiation, a way to assess how the scatter component varies with energy is a MC approach.

**MC simulation of SPECT of Y-90 Bremsstrahlung radiation**

Previous works have been done at this regard, using different software and methodologies **[5, 6]**. One work, using a MC simulation and based on scatter removal and a resolution modeling of the camera, allowed improvement in the contrast of hot and cold region **[7]**.

In this work, we used the Gate (Geant4 Application for Tomographic Emission, **[3]**) MC code. Gate is an application for tomographic emission imaging using the Geant4 toolkit to model the interaction of particles with matter **[8]**. A substantial part of our work consisted of testing and analyzing the capabilities of the MC code in simulating the decay kinetic and the SPECT head geometry, by using two different models for the EM Physics, in order to find the best setup to reproduce the real experiment: the Penelope-based low energy model **[9]**, and the Standard model **[10]**. The results of this analysis led us to use, for our purpose, the Standard model, focusing appropriate electromagnetic processes on those covered by our study.

To simulate the liver focality in the patients body (the average volume of HCC tumors for n=59 patients treated in our center is 184 cc (range 3-904 cc), a *Jaszczak* phantom with a smaller cylindrical hot volume (166 cc) with Y-90 microspheres was modeled with the MC and a uniform random generation of $\beta^-$ particles with energies distributed according the Y-90 decay spectrum **[11]** were used.

The decomposition of the Bremsstrahlung radiation emitted by the phantom (**Fig.1**) shows that the clinical image is mainly due to scattered photons: primary photons amount to about 30% of total counts. This result seems reasonable if we consider that *Minarik, D. et al.* **[12]** showed that, in a *Petri* dish, an extremely small volume to favor scattering, primary and non-primary photons are already quantitatively comparable.

The normalized primary component to the total MC spectrum (i.e. the normalized difference between the total MC spectrum (MCStd) and its non-primary (NP) component) as a function of energy, shows a minimum around 47 keV, where primary photons amount to ~20% of all detected events. Starting from this energy, residuals have a slowly increasing trend before reaching an extended similar-plateau at about 250 keV, where primary photons amount to ~35% of all detected events (**Fig.2**)

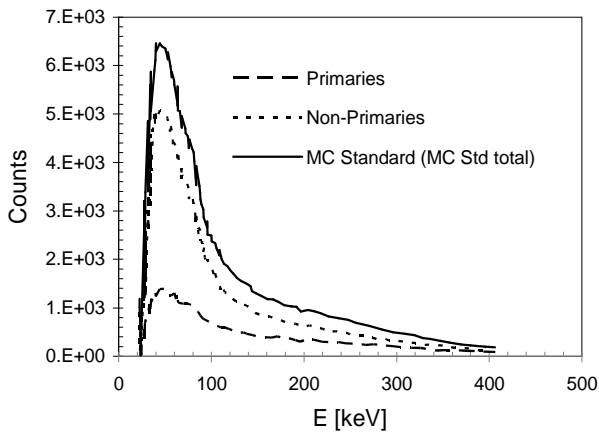
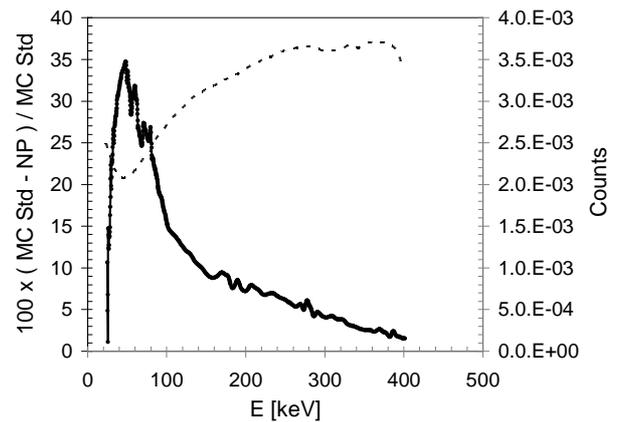

**Fig.1** MC simulated photon spectrum emitted by the Y-90 cylindrical phantom as detected by the gamma-camera. Primary photons (dashed line) and non-primaries (dotted line).

**Fig.2** Normalized primary component to the total MC spectrum (dashed line, left vertical axis). Superimposed, the primary component (solid line, right vertical axis).

We recall that our simulation suffers from some limitations, mainly due to the limited information available on the detector datasheets (which resulted in a non-optimal reproduction of the backscattering region of the gamma-camera head), that can affect the shape of the curve reported in **Fig.2**. For these reasons we used the MC outcomes as "a guide" to have an indication of the region where the primary fraction is higher, which resulted to be the right part of the spectrum.

**Linear Source measurements**

The two most important effects that degrade the accuracy and quality of nuclear medicine imaging are photon attenuation and scattering, resulting from photon interactions with tissues in the patient's body. These effects are more pronounced for photons of high energy. Therefore, the efficiency of the detector used is closely linked to the photon energy. For example, NaI crystals used in SPECT cameras provide high (close to 80 - 90%) sensitivity, but only for the detection of relatively low energy 140 keV photons; this sensitivity decreases to only 25 - 28% when 511 keV photons are involved **[13]**. Indeed, its absorption coefficient decreases rapidly with increasing of energy. From about 500 keV, representing the maximum detectable energy of the gamma camera, there is a rapid decrease of the NaI(Tl) absorption coefficient when the photon energy increases. This is the main factor of the efficiency decrease of the detector. The intrinsic efficiency $\varepsilon$ behavior of NaI(Tl) versus the photon energy for different detector thickness is given by: $\varepsilon = 1 - e^{(-\mu_\ell(E) \cdot x)}$, where $x$ is the detector thickness and $\mu_\ell(E)$ its linear attenuation coefficient at the photon energy $E$, which increases linearly with absorber density and reflects the "absorptivity" of the absorbing material.

Referring to Fig.2, the higher energy portion of the Bremsstrahlung spectrum seems to be characterized by the higher fraction of primary photons. On the other hand, with the increase of the energy, the probability of penetrating the septa of collimation becomes higher.

Our idea was to measure the LSF of a linear Y-90 inserted in a scattering medium, to see if there would be some indication about the influence of these two opposing effects on the LSF shape. We acquired a linear source of Y-90 that was inserted into the central hole of the CT body and head phantoms, with the expectation of finding a minimum of LSF. As already pointed, although these phantoms do not represent a realistic scattering material surrounding the liver tumor, the large amount of thickness around the Y-90 linear source has the effect of maximizing the quantity of scattering radiation, amplifying its influence on LSF shape. The Y-90 linear source (approximately 50 MBq) was prepared using a glass capillary tube with a hole of ∅ 1 mm, which was first surrounded by a thickness of Nylon (CT number = 60) and then was inserted in the central hole of the CT PMMA phantoms. Emission data were acquired using a tomographic protocol with six simultaneous and contiguous energy windows of [35-59], [60-84], [85-114], [115-165], [165-215] and [215-279] keV and the following parameters:

1. SPECT: Matrix128x128 (pixel size = 4.8 mm) – Number of projections = 30, 120 seconds/projection – Circular orbit (R = 30cm) – HE collimators;

2. CT: Spiral mode – 130 kV – 0.8 s/rotation – 61 slices 5.0 mm thick – Pitch 1.4.

The reconstruction parameters were the following:

3. SPECT: Iterative 3D OSEM "Flash 3D" – Subsets = 6 ; Iterations = 4 (4i6s) – Gaussian Filter : FWMH = 8.4 mm – Attenuation Correction (AC) applied;

4. CT: Kernel 1: B08s (SPECT AC) – Kernel 2: B41s (medium) – FOV: 500 mm – Slice thickness = 5.0 mm – Reconstruction increment = 2.5mm .

The SPECT-CT protocol performs an attenuation correction on the acquired events based on the CT dataset. We applied the correction based on the central value of each energy window.

**Contrast and CRC measurements with IEC NEMA phantom**

The NEMA IEC phantom (mod. PET/IEC-Body/P) for the study of the image quality in PET filled with Y-90 chloride was acquired with the SPECT-CT system in order to assess the Contrast and the CRC using different energies and setups. Y-90 chloride was preferred to Y-90 microspheres, because the latter suddenly fell down to the bottom of the phantom compartments (spheres and background) making it impossible to obtain homogeneous solutions. A vial of 1.41 ml with a total activity of 3071 MBq at the calibration time was used. A total of eight SPECT-CT acquisitions were performed: four acquisitions with Y-90 activity only in the spheres, four acquisitions with Y-90 activity also in the background compartments. **Tabs. 1** and **2** summarize the Y-90 activity concentrations for the eight acquisitions.

| SET | $(A_S(t))/V_S$ (MBq/mL) |
|---|---|
|  | (Spheres) |
| 1$^{th}$ acquisition | 3.374 |
| 2$^{nd}$ acquisition | 3.322 |
| 3$^{th}$ acquisition | 2.627 |
| 4$^{th}$ acquisition | 2.586 |

**Tab.1** Activity concentration in the six spheres for the first set of 4 acquisitions, without activity in the background compartment ($A_S$ = activity in the spheres, $V_S$ = volume of spheres).

| SET | $(A_S(t))/V_S$ (MBq/mL) | $(A_B(t))/V_B$ (MBq/mL) |
|---|---|---|
|  | (Spheres) | (Background) |
| **5$^{th}$ acquisition** | 1.192 | 0.153 |
| **6$^{th}$ acquisition** | 1.167 | 0.150 |
| **7$^{th}$ acquisition** | 0.920 | 0.118 |
| **8$^{th}$ acquisition** | 0.695 | 0.089 |

**Tab.2** Activity concentration in the six spheres and in the background compartment for the second set of 4 acquisitions ($A_S$ = activity in the spheres, $V_S$ = volume of spheres, $A_B$ = total activity in the background, $V_B$ = volume of background)

Based on the results provided by the measurements on the Y-90 line source, which suggested the energy range of 100 - 140 keV as a good candidate to meet the minimization of scattering and septa penetration effect, we designed two different acquisition windows-sets: a narrow window-set to cover the region 50 – 196 keV and a large window-set to cover the region 40 – 440 keV to extend the analysis to the higher portion of the spectrum. The *narrow* window-set was not centered exactly on 120 keV, but it was shifted toward lower energies to include the peak of characteristic X-rays emitted by the lead collimator, in order to analyze their effect on the image quality.

Each of these two window-sets is composed of six contiguous sub-windows, which are reported in **Tab.3** and **Tab.4**. In **Fig.3** we report the acquisition windows superimposed to the Bremsstrahlung spectrum acquired by the SPECT detectors equipped with lead HE collimators. The peak around 73 - 87 keV is due to the characteristic X-rays emitted by the lead.

| Sub-window | 1$^{th}$ | 2$^{nd}$ | 3$^{th}$ | 4$^{th}$ | 5$^{th}$ | 6$^{th}$ |
|---|---|---|---|---|---|---|
| Size (keV) | 59.5 | 58.5 | 58.9 | 60.0 | 80.0 | 80.0 |
| Center (keV) | 70.0 | 130.0 | 190.0 | 250.0 | 320.0 | 400.0 |

**Tab.3** Size and central position in energy [keV] for the sub-windows of the *large* window-set.

| Sub-Window | 1$^{th}$ | 2$^{nd}$ | 3$^{th}$ | 4$^{th}$ | 5$^{th}$ | 6$^{th}$ |
|---|---|---|---|---|---|---|
| Size (keV) | 24.8 | 23.2 | 24.2 | 24.3 | 24.0 | 23.9 |
| Center (keV) | 62.0 | 86.0 | 110.0 | 135.0 | 160.0 | 184.0 |

**Tab.4** Size and central position in energy [keV] for the sub-windows of the *narrow* window-set

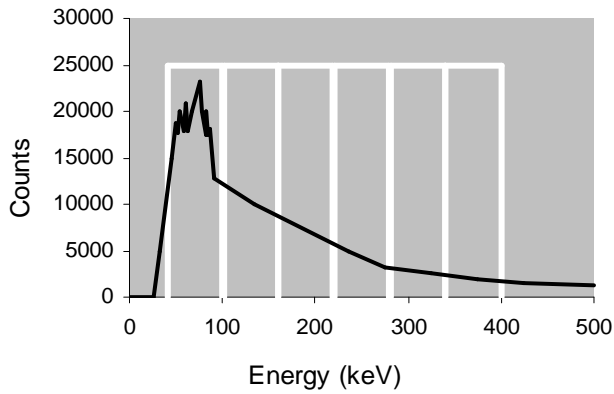 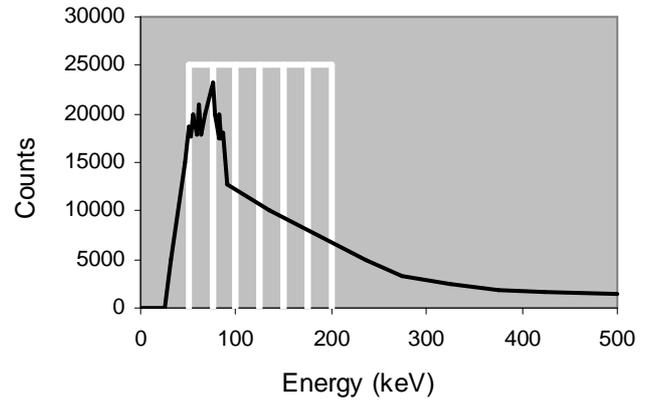

Fig.3a                                    Fig.3b

**Fig.3** *Large (a)* and *narrow* (b) window-set with superimposed the Bremsstrahlung photon spectrum measured by the gamma-camera with HE collimators.

In **Tab.5** we summarize the acquisition parameters and setups for the eight SPECT-CT acquisitions on the IEC-NEMA phantom. All the acquisitions were made using a matrix of 256x256 pixels, except one of 128x128 pixels. Six acquisitions were made using HE collimators, except one with ME and one with LE collimators.

| SET | Windows | Collimator | Orbit | Matrix | Time (s/view) |
|---|---|---|---|---|---|
| 1$^{th}$ acquisition | narrow | HE | circular | 256 x 256 | 180 |
| 2$^{nd}$ acquisition | large | HE | circular | 256 x 256 | 120 |
| 3$^{th}$ acquisition | narrow | HE | non circular | 256 x 256 | 180 |
| 4$^{th}$ acquisition | large | HE | non circular | 256 x 256 | 120 |
| 5$^{th}$ acquisition | large | HE | non circular | 256 x 256 | 240 |
| 6$^{th}$ acquisition | large | HE | non circular | 128 x 128 | 240 |
| 7$^{th}$ acquisition | large | ME | non circular | 256 x 256 | 360 |
| 8$^{th}$ acquisition | large | LEHR | non circular | 256 x 256 | 360 |

**Tab.5** Configuration and parameters of the SPECT-CT acquisition on the NEMA-IEC phantom.

Count rates were always < 2.5 kcps i.e. within the linear range of the SPECT system.
To assess the Contrast and the CRC we followed the procedure described in **[1]**. For each reconstruction dataset, we identified the transversal image containing the centers of the spheres and drew circular Regions Of Interest (ROI) guided by the CT images, in order to calculate the average intensity ($C_S$). On the same planes, a ROI was also drawn around the background compartment boundary, and the average intensity was calculated ($C_B$). Contrast and CRC and their standard deviations were calculated according with the formulas summarized in **Tab.6**. The reconstruction parameters were the same as used for the linear source measurements, i.e. 3D OSEM "Flash 3D with 4 iterations and 6 subsets (4i6s). Furthermore, in order to investigate any possible influence given by the number or iterations and subset, the data were also reconstructed using the reconstruction settings 2i-4s and 6i-10s.

The SPECT-CT protocol performs an attenuation correction on the acquired events based on the CT dataset. We applied the correction based on the central value of each energy window

| Parameter | Formula | Standard deviation | definitions |
|---|---|---|---|
| Contrast | $C_l = \dfrac{C_S - C_B}{C_B}$ $= \dfrac{C_S}{C_B} - 1$ | $\sigma_{C_l} = \dfrac{1}{C_B}\sqrt{\sigma_{C_S}^2 + \left(\dfrac{C_S}{C_B}\right)^2 \sigma_{C_B}^2}$ | $\sigma_{B_C} = \sqrt{B_C}$ $\sigma_{S_C} = \sqrt{S_C}$ $\sigma_{C_B} = \dfrac{\sqrt{B_C}}{V_B}$ $\sigma_{C_S} = \dfrac{\sqrt{S_C}}{V_S}$ |
| CRC | $CRC = \dfrac{\dfrac{C_S}{C_B} - 1}{R - 1}$ | $\sigma_{CRC} = \dfrac{1}{C_B(R-1)}\sqrt{\sigma_{C_S}^2 + \left(\dfrac{C_S}{C_B}\right)^2 \sigma_{C_B}^2}$ | |

$B_C$ = total counts in the background compartment
$V_B$ = background volume in number of voxels
$C_B = B_C/V_B$ = number of counts in the background compartment per voxel
$S_C$ = total counts in the lesion
$V_S$ = lesion volume in number of voxels
$C_S = S_C/V_S$ = number of counts in the lesion (spheres) per voxel
R = true sphere-to-background activity concentration ratio

**Tab.6** Expressions used in the calculation of Contrast and CRC.

**Discussion and Results**

**MC outcomes**

According to the theory of $\beta$ decay, the histogram reported in **Fig.2** shows that the unscattered photons are most numerous at low energies. This matches the result found by **[5]**. A comparison with the scattered component is however more indicative for our purposes. **Fig.2** tells also us that the fraction of primary detected photons (i.e. photons that have not undergone Compton scatter) increases with energy with respect to the scattered component, always being the smallest component of the total image counts. *Heard, S. et al.***[5]** found that, as energy increases with MEGP, and for all energies with LEGP collimators, camera effects (collimator scatter and septa penetration, Pb X-rays and backscatter) dominate the total image counts. Taking into account these effects, they found an optimal image Contrast for both LEGP and MEGP collimation in the region 100 - 150 keV.
As already pointed out, our MC simulation suffers from a non-optimal reproduction of the region of the photomultipliers and electronics interfaced to the scintillator. Taking however these limitations into account, the MC outcomes suggest that in order to reduce the scattering fraction in the images, we should place the acquisition window on the right side of the spectrum. On the other hand, with the increasing of their energy, photons penetrate more both the collimator septa and also the detector, with consequent degradation of the image quality (contrast and spatial resolution), and decrease in intensity. For these reasons, the optimization of the window acquisition position is of great importance.
Regarding septa penetration, following the equations in **[14]**, the path length *w* for a photon travelling across two collimator's holes is related to the septa thickness *t*, to the hole diameter *d* and to the hole lengthy *l* by:

$$t \approx \frac{2dw}{l - w} \quad (3)$$

( $t$ = 2mm, $d$ = 0.4 cm, $l$ = 5.97 cm). The septa penetration fraction is simply $e^{-w\mu}$, being $\mu = \mu(E)$ [cm$^{-1}$] the linear attenuation coefficient for lead.

From 200 keV septa penetration rapidly increases with energy, and above 410 keV, the thickness of the septa actually is not sufficient to limit to 5% the fraction of photons that pass through them.

## Line Spread Function (LSF) of the Y-90 linear source

To estimate qualitatively the influence of the scattering component and the septa penetration effect along the energy axis, we analyzed the activity profiles of the Y-90 linear source for each of the six tomographic datasets acquired within the 6 energy windows centered on 47, 72, 100, 140, 190 and 247 keV. For each dataset, all the axial reconstructed images (n = 62) were added together in order to increase the counting statistics. **Fig.4a** shows as example the activity profiles for the first window, normalized in area, for the head and body CT phantoms.

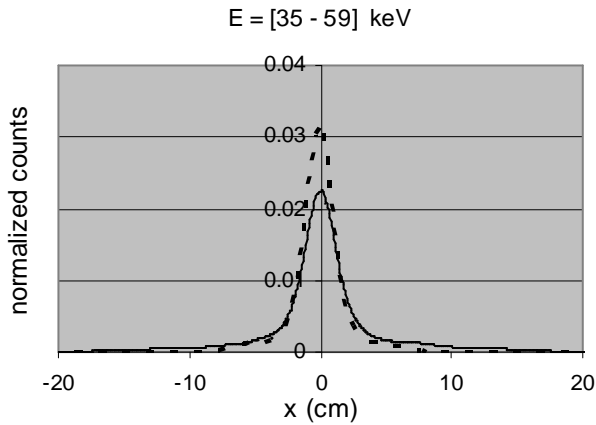

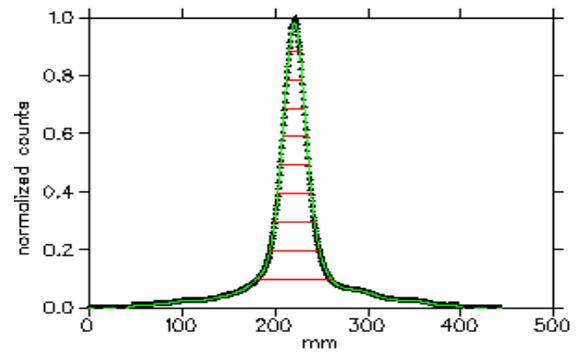

**Fig. 4a** Activity profiles for the first energy window, normalized in area, for the head (dashed line) and body (continuous line) CT phantoms.

**Fig. 4b** LSF widths at different percentage of the maximum.

To characterize the activity profiles, the FWHM is not sufficient because of the presence of tails at the sides of the LSFs, which contributes to the image degradation, and therefore also the FWTM has to be assessed. Furthermore, since in the clinical image post-processing a reasonable value to be applied for the counts threshold is around [35-55] % **[15]**, we used as figure of merit what we called *FW@n%M*, namely the Full Width at different percentage (*n %*) of the Maximum of the activity profiles (**Fig.4b**). The values of the *FW@n%M* are reported in **Figs.5** for the LSFs normalized at their maximum value. The behaviors along the energy do not change if the activity profiles are normalized in area.

Looking at the plots, there is an evidence of a minimum around ~140 keV, even more pronounced for n < 50-60, in particular for n = 10, which corresponds to the FWTM.

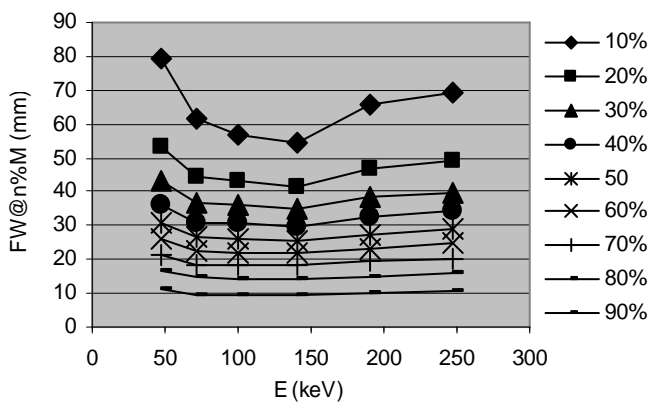

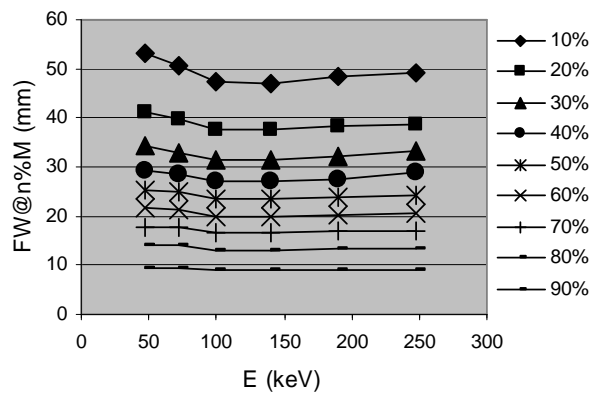

**Fig.5a** Full widths at *n*% of maximum vs. energy for the body phantom.

**Fig.5b** Full widths at *n*% of maximum vs. energy for the head phantom.

**Contrast and Contrast Recovery Coefficient**

First we discuss the results for the Contrast measured on IEC-NEMA phantom, obtained with the first four acquisitions, with activity in the six spheres and without activity in the background compartment (**Tab.1**).

We report in **Fig.6** the behaviors of the Contrast with the energy for each sphere. **Fig.6b** shows the Contrast measured with the HE collimators in the energy interval of 40 – 440 keV (*large* window-set), where a maximum is recognizable, for the three largest spheres, at 130 keV which is the center of the second sub-window (~101 - 159 keV). **Fig.6b** refers to an acquisition with a circular orbit; **Fig.6d** refers to an acquisition with a non-circular orbit, i.e. with a movement of the gamma-camera head that follows the patient contour.

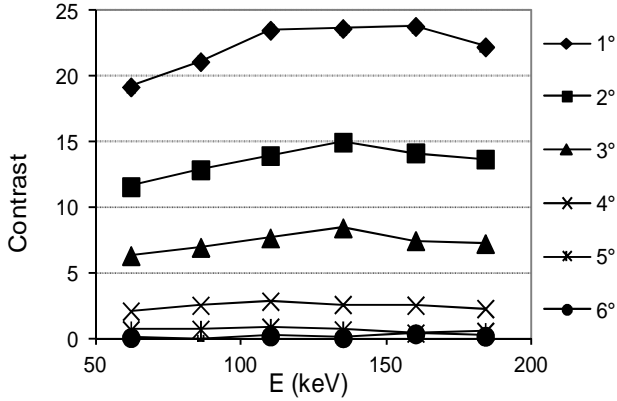

**Fig.6a** Contrast of the six spheres - 1$^{th}$ acquisition, 256x256 matrix, HE collimator, circular orbit (see **Tab.5**).

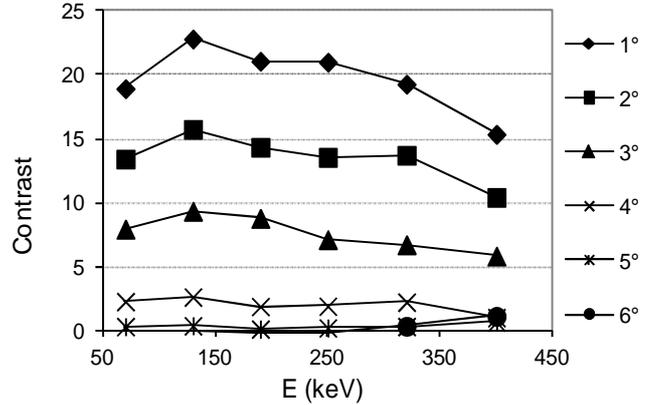

**Fig.6b** Contrast of the six spheres – 2$^{nd}$ acquisition, 256x256 matrix, HE collimator, circular orbit (see **Tab.5**).

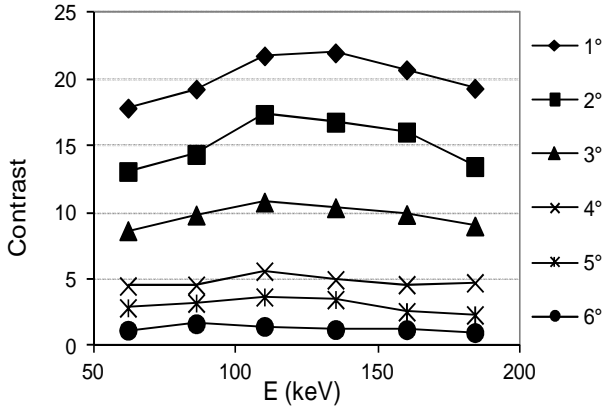

**Fig.6c** Contrast of the six spheres - 3$^{th}$ acquisition, 256x256 matrix, HE collimator, non circular orbit (see **Tab.5**).

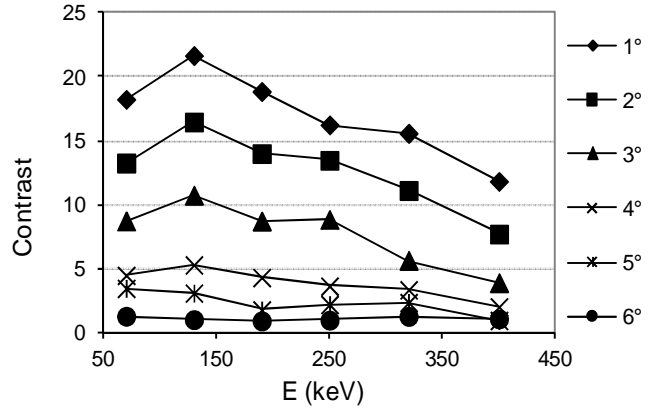

**Fig.6d** Contrast of the six spheres - 4$^{th}$ acquisition, 256x256 matrix, HE collimator, non circular orbit (see **Tab.5**).

**Fig.6a** and **Fig.6c** show the Contrast with a higher energy resolution within the interval of 50 – 196 keV (*narrow* window-set), and show that the maximum is placed between the centers of the third and fourth sub-windows, i.e. between 110 and 135 keV.

The second set of four SPECT-CT acquisitions (**Tab.2**) was carried out with activity in the background compartment. **Fig.7** reports the behaviors of the CRC with the energy for each sphere. The fifth and the sixth acquisition (**Figs.7a** and **7b**), both performed with HE collimators and with the *large* window-set of 40 – 440 keV, can be compared to highlight possible differences of CRC due to the different acquisition matrix, which was 256x256 for the fifth and 128x128 for the sixth acquisition. It can be observed that, as expected and a part of slight fluctuations, the CRC trends are comparable for both, and a maximum is recognizable, for three largest spheres, at 130 keV which is the center of the second sub-window (101-159 keV).

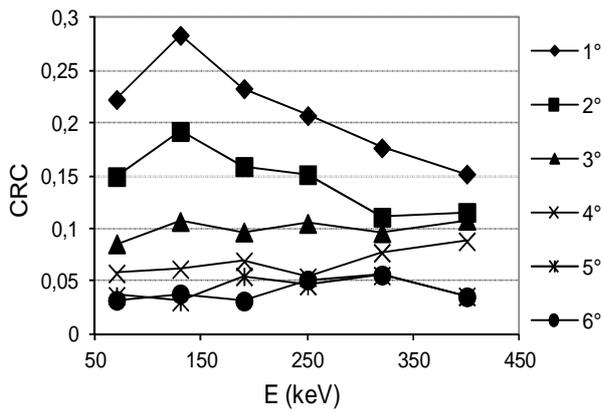
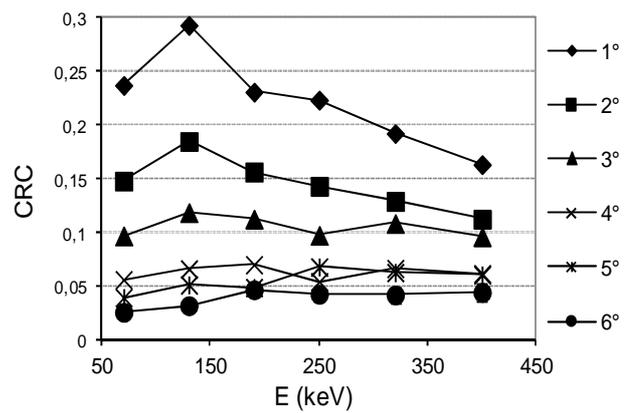

**Fig.7a** Contrast Recovery Coefficient of the six spheres - 5[th] acquisition, 256x256 matrix, HE collimator, non circular orbit.

**Fig.7b** Contrast Recovery Coefficient of the six spheres - 6[th] acquisition, 128x128 matrix, HE collimator, non circular orbit.

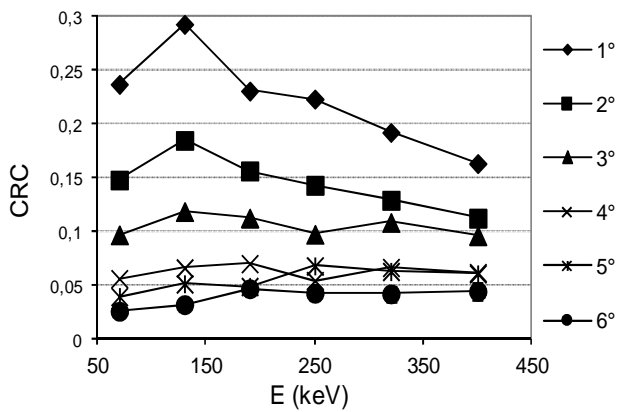
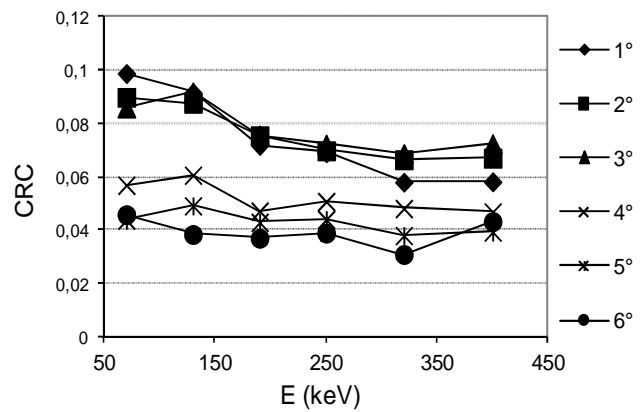

**Fig.7c** Contrast Recovery Coefficient of the six spheres – 7[th] acquisition, 256x256 matrix, ME collimator, non circular orbit.

**Fig.7d** Contrast Recovery Coefficient of the six spheres - 8[th] acquisition, 256x256 matrix, LEHR collimator, non circular orbit.

It is noteworthy at this point to emphasize the agreement between the energies which were found at the minimum of the *FW@n%M* (~140 keV) for the LSF of the linear Y-90 source and those that maximize the Contrast and the CRC (~ 110-135 keV). These results are in agreement with **[16]**, where the authors claim an optimal energy window of 100 - 160 keV, using a completely different figure of merit.

The purpose of the last two acquisitions (**Tab.5**) was to study the effects of the collimator type on the CRC. **Fig.7d** shows that the maximum value of the CRC for LEHR collimator is about 1/3 compared to HE collimator, and the trend is slowly decreasing with energy without any local maximum. Instead, the shape of CRC measured with the ME (**Fig.7c**) and HE (**Fig.7a**) collimators is approximately the same, again with the presence of a maximum at 130 keV (more recognizable for the two largest spheres), but with lower CRC values for ME collimators compared to HE collimator. In this respect, in **Fig.8** we report the Contrast calculated for the largest sphere relatively to the fifth, seventh and eighth acquisitions: the Contrast is higher when the images are acquired with HE collimators; this result is not surprising since the Contrast and CRC are related by the (R-1) constant.

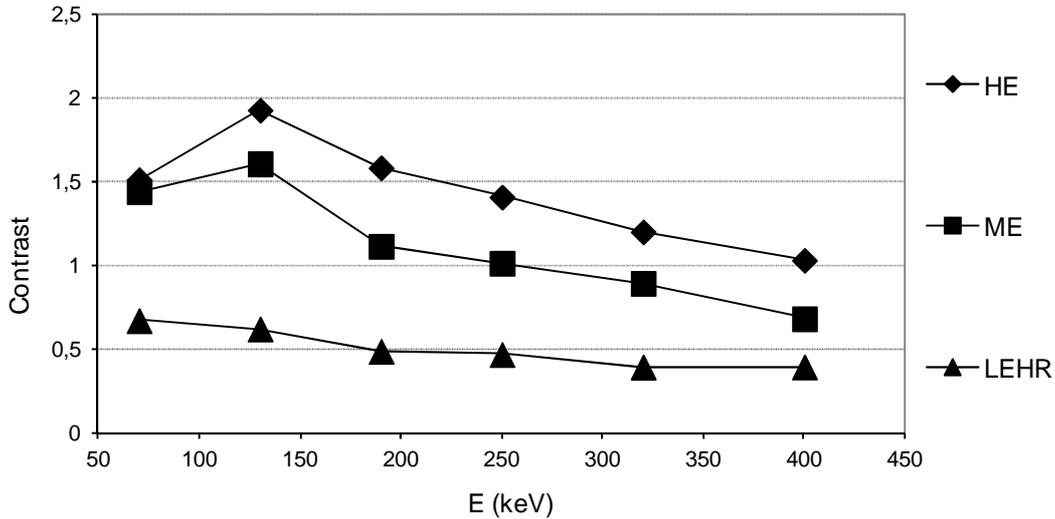

**Fig.8** Contrast for the biggest sphere in the 5[th] acquisition (HE collimator), 7[th] acquisition (ME collimator) and 8[th] acquisition (LEHR collimator).

In addition to the 4i6s reconstruction set, we also investigated the energy dependence of the contrast and the CRC for the two different sets 2i4s and 6i10s. The 2i4s combination has a lower number of iterations and subsets, and leads to lower values of contrast and CRC. Conversely, the 6i10s combination leads to a increasing of contrast and CRC, but at the expense of a longer reconstruction time. These results are in agreement with **[17]**, where the authors found that increasing the product between the iteration's number and subset's number in OSEM resulted in improved contrast, but at the expense of a increase of the noise. Anyhow the overall behaviour of the contrast and the CRC along the energy reported in **fig.6** and **fig.7** is preserved with the presence of a maximum at 110-135 keV for the larger spheres. By way of example we report in **fig.9** the CRC measured along the energy for the 5[th] acquisition of tab.5 with the 2i4s (**fig.9a**) and 6i10s (**fig.9b**) reconstruction sets.

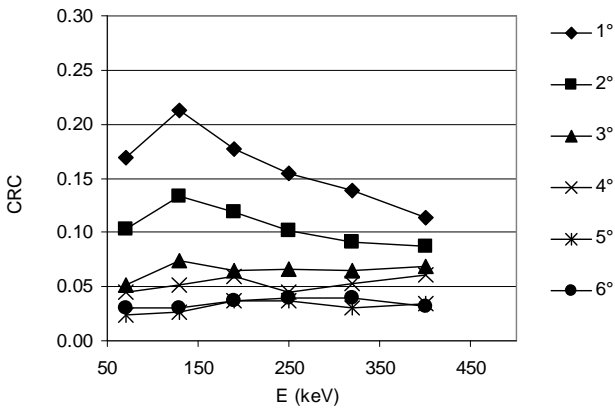
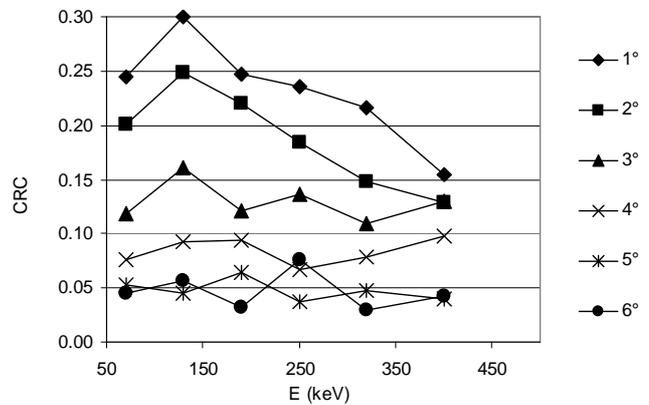

**Fig9a** Contrast Recovery Coefficient of the six spheres - 5[th] acquisition of tab.5, 256x256 matrix, HE collimator, non circular orbit, 2i4s reconstruction set

**Fig.9b** Contrast Recovery Coefficient of the six spheres - 5[th] acquisition of tab.5, 256x256 matrix, HE collimator, non circular orbit, 6i10s reconstruction set

In tab.7a we report the contrast values for the three larger spheres measured at 135 keV for the 3 combinations 2i4s, 4i6s and 6i10s, in correspondence of the 1[th] acquisition of tab.5. In tab.7b we report the CRC values for the three larger spheres measured at 130 keV for the 3 combinations 2i4s, 4i6s and 6i10s, in correspondence of the 5[th] acquisition of tab.5.

| Reconstruction set | Contrast | | |
|---|---|---|---|
| | sphere ⌀ 3.7 cm | sphere ⌀ 2.8 cm | sphere ⌀ 2.2 cm |
| 2i4s | 14.7 | 6.8 | 2.7 |
| 4i6s | 23.7 | 15.0 | 8.5 |
| 6i10s | 28.7 | 22.1 | 13.8 |
| Tab.7a | | | |

| Reconstruction set | CRC | | |
|---|---|---|---|
| | sphere ⌀ 3.7 cm | sphere ⌀ 3.7 cm | sphere ⌀ 3.7 cm |
| 2i4s | 0.21 | 0.21 | 0.21 |
| 4i6s | 0.28 | 0.28 | 0.28 |
| 6i10s | 0.30 | 0.30 | 0.30 |
| Tab.7b | | | |

Tab. 7a : contrast measured al 135 keV (1[th] acquisition of tab.5) of the three larger spheres, for the three different reconstruction sets. Tab. 7b : CRC measured al 135 keV (5[th] acquisition of tab.5) of the three larger spheres, for the three different reconstruction sets (i = iterations s = subset).

**Conclusions**

We presented some new results about optimization of SPECT/CT acquisition of Bremsstrahlung imaging for liver cancer therapies with Y-90 microspheres.

The optimization is important to improve the image quality, which is also reflected on the patient image-based dosimetry. It is well known that the quality of Bremsstrahlung images for patients treated with Y-90 is poor, because of scattered radiation and collimator septa penetration.

In this work we have had first a theoretical approach, simulating with a MC (Geant4/Gate) a SPECT acquisition of a cylindrical phantom with Y-90. MC outcomes suggested that we perform an experiment with a linear source of Y-90 inserted in a scattering material with the purpose of identifying the energy that could be interpreted as the compromise between the minimization of the scattering component and the septa penetration effect.

In order to improve the image quality and the quantitative imaging, different quantities need to be analyzed: in this work we investigated how the Contrast and the CRC vary with photon energy as selected by the acquisition window.

Our results showed that that Contrast and CRC are higher for HE collimators compared to ME collimators (lead). For the large window the results shown that the they have a net maximum between ~ 70-190 keV for the larger spheres, which is restricted between ~ 110-135 keV with the narrow window-set, leaving some flexibility to choose the optimum acquisition window size, depending on the required counting statistic. In the light of the presented results, we are currently using an energy window between 100 - 200 keV, divided into five equals sub-windows, each one of 20 keV size, in order to have a more accurate attenuation correction. The choice to extend the upper limit of up to 200 keV is due increasing the statistics. Future work will involve the study of other image quality parameters such as the Contrast to Noise Ratio and Spatial Resolution.

**Acknowledgments**

They also thank the radiographers Livio Bastianutti, Diego Primossi, and Davide Stanic for their help in carrying out the measurements with the SPECT-CT equipment and Mr. Sergio Spivac for the processing and modification of the glass capillary. A special thank to SIRTEX ® Company, in particular Dr. Michael Tapner, which provided the Y-90 chloride to carry out the measurements.

**Conflict of interest**

None.